\documentstyle[12pt,epsfig]{article}
\newfont{\sfb}{cmssbx10 scaled 1400}
\newfont{\bigsf}{cmssbx10 scaled 1600}

\begin{document}
\thispagestyle{empty}
\begin{flushright}
KOBE--FHD--02--03\\
January ~~~2002
\end{flushright}
\begin{center}
{\baselineskip 22pt
{\LARGE\bf Semi-inclusive large--$p_T$ light hadron pair production
as a probe of polarized gluons\footnote[2]{Talk given by T. Morii at the EMI2001 
(International Symposium on Electromagnetic Interactions in Nuclear and Hadron Physics), 
RCNP, Osaka, Dec. 4--7, 2001.}\\
}}
\vspace{2.0em}

T. Morii\\
\vspace{0.8em}
{\it Faculty of Human Development and}\\
{\it Graduate School of Science and Technology,}\\
{\it Kobe University, Nada, Kobe 657--8501, Japan}\\
E--mail: {\tt morii@kobe-u.ac.jp}

\vspace{1em}
Yu-Bing Dong\\
\vspace{0.8em}
{\it Institute of High Energy Physics, Academia Sinica,}\\
{\it Beijing 100039, P. R. China}\\
E--mail: {\tt dongyb@mail.ihep.ac.cn}

\vspace{1em}
and\\
\vspace{1em}
T. Yamanishi\\
\vspace{0.8em}
{\it Department of Management Science,}\\
{\it Fukui University of Technology, Gakuen, Fukui 910--8505, Japan}\\
E--mail: {\tt yamanisi@ccmails.fukui-ut.ac.jp}

\vspace{2cm}
{\bf Abstract}
\end{center}

\baselineskip=22pt

We propose a new formula for extracting the polarized gluon 
distribution $\Delta g$ in the proton from the 
large--$p_T$ light hadron pair production in deep inelastic
semi-inclusive reactions.  Though the process dominantly
occurs via photon--gluon fusion(PGF) and QCD Compton, 
we can remove an effect of QCD Compton from the combined
cross section of light hadron pair production by using symmetry relation 
among fragmentation functions and thus, rather clearly extract 
information of $\Delta g$.

\clearpage

\section{Introduction}

Since the measurement of the polarized structure 
function of the proton $g_1^p$ for wide kinematic range of 
Bjorken $x$ and $Q^2$ by the EMC in 1988\cite{EMC88},
the so-called proton spin puzzle has been still one of the most
challenging topics in particle and nuclear physics.
After the EMC experiment, much progress has been attained experimentally
and theoretically.\cite{Review}  
A large amount of data on polarized structure
functions of proton, neutron and deuteron were accumulated.  The
progress in the data precision is also remarkable.  
Furthermore, with much theoretical effort based on the 
next--to--leading order QCD analyses of
the longitudinally polarized structure function $g_1$,
a number of excellent parameterization models of the
polarized parton distribution functions (pol--PDFs) were obtained 
from fitting to many/precise data with various targets for
polarized deep inelastic scattering (pol--DIS)
\cite{GRSV96,GS96,polPDF,Goto}.
All of these parameterization models
tell us that quarks carry a little of the proton spin, i.e. 
only 30\% or smaller.  Due to an intense experimental and theoretical
effort, a rather good knowledge of the polarized valence
quark distribution has been obtained so far.
However, knowledge of $\Delta g$ is still very limited.
As is well-known, the next--to--leading order QCD analyses on $g_1$
bring about information on the first moment of the polarized
gluon $\Delta g$.\cite{Mertig}
However, there are large uncertainties in $\Delta g$ extracted
from $g_1$ alone.
To solve the so-called spin puzzle, it is very important
to precisely know how the gluon polarizes in the nucleon.

So far, a number of interesting processes such as
direct prompt photon production in polarized proton--polarized proton
collisions\cite{Craigie}, open charm\cite{Watson} and
$J/\psi$\cite{Morii} productions
in polarized lepton scattering off polarized nucleon targets,
were proposed for studying longitudinally
polarized gluon distributions.
Recently, HERMES group at DESY reported the first
measurement of the polarized gluon distribution from
di--jet analysis of semi--inclusive processes in
pol--DIS, though only one data point was given as a function
of Bjorken $x$.\cite{HERMES}
In general, a large--$p_T$ hadron pair is produced via
photon--gluon fusion (PGF) and QCD Compton at the lowest order
of QCD.
The PGF gives us a direct information on the
polarized gluon distribution in the nucleon, whereas QCD Compton
is background to the signal process for extracting 
$\Delta g$.\cite{Bravar,Roeck}
For the case of heavy hadron pair($D\bar D$, $D^*\bar D^*$, etc.) 
production, we could safely neglect the contribution of QCD Compton 
because the heavy quark content in the proton 
is extremely small and also the fragmenting 
probability of light quarks to heavy hadrons is very small.
However, in this case the cross section itself
is small at the energy in the running(HERMES) or
forthcoming(COMPASS) experiments and thus, we could not have enough
data for carrying out detailed analysis.  For the case of light
hadron pair production, we will have a rather large number of events
because of large cross sections.  
However, in this case QCD Compton can no longer be neglected and hence,
it looks rather difficult to unambiguously extract the behavior
of $\Delta g$ from those processes.

In this work, to overcome these difficulties 
we propose a new formula for clearly extracting 
the polarized gluon distribution from the light hadron pair 
production of pol--DIS by removing the QCD Compton component 
from the cross section.
As is well-known, the cross section of the hadron pair production
being semi-inclusive process, can be calculated based on the parton model
with various fragmentation functions.  Then, by using symmetry relations
among fragmentation functions and taking an appropriate combination
of various hadron pair production processes, we can remove the
contribution of QCD Compton from the cross section
and thus, get clear information
of $\Delta g$ from those processes.
Here, to show how to do this practically, we consider the light hadron pair
production with large transverse
momentum.

\section{A new formula for extracting $\Delta g(x)$ from 
large--$p_T$ pion pair production}

Let us consider the process of  $\ell + N \to \ell' + h_1 + h_2 +X$
in polarized lepton scattering off polarized nucleon targets, 
where $h_1$ and $h_2$ denote light hadrons in a pair.
As mentioned above,
the spin--dependent differential cross section at the leading order of
QCD can be given by the sum of the
PGF process and QCD Compton as follows,
\begin{equation}
d\Delta\sigma = d\Delta\sigma_{PGF} + d\Delta\sigma_{QCD}~,
\label{eqn:2}
\end{equation}
with
$$
d\Delta\sigma=d\sigma_{-+}-d\sigma_{++}+d\sigma_{+-}-d\sigma_{++}~,
$$
where, for example, $d\sigma_{-+}$ denote that the helicity of
an initial lepton and the one of a target proton is
negative and positive, respectively.
Each term on the right hand side of eq.(\ref{eqn:2}) is given by
\begin{eqnarray}
&&d\Delta\sigma_{PGF}\nonumber\\
&&\hspace{.5cm} \sim  \Delta g(\eta, Q^2)
d\Delta\hat\sigma_{PGF}\hspace*{-.3cm}
 \sum_{i=u, d, s, \bar u, \bar d, \bar s} \hspace*{-.4cm}
e_i^2 \{ D^{h_1}_i(z_1, Q^2) D^{h_2}_{\bar i}(z_2, Q^2) 
+(1\leftrightarrow 2)\},\\
\label{eqn:3}
&&d\Delta\sigma_{QCD}\nonumber\\
&&\hspace{.5cm} \sim \hspace*{-.3cm}
 \sum_{q=u, d, s, \bar u, \bar d, \bar s} \hspace*{-.4cm}
e_q^2 \Delta f_q(\eta, Q^2) d\Delta\hat\sigma_{QCD}
\{ D^{h_1}_q(z_1, Q^2) D^{h_2}_g(z_2, Q^2) 
+(1\leftrightarrow 2 )\},
\label{eqn:4}
\end{eqnarray}
where $\Delta g(\eta, Q^2)$, $\Delta f_q(\eta, Q^2)$ and
$D^h_i(z, Q^2)$ denote the polarized gluon and the quark distribution
function of $q$ with momentum fraction $\eta$ and the fragmentation
function of a hadron $h_i$ with momentum fraction $z_i$ emitted
from a parton $i$, respectively.
$d\Delta\hat\sigma_{PGF}$ and $d\Delta\hat\sigma_{QCD}$ are the
polarized differential cross sections of hard scattering subprocesses for
$\ell g\to \ell' q\bar q$ and 
$\ell \stackrel{\scriptscriptstyle(-)}{q}\to \ell' \stackrel{}{g}
\stackrel{\scriptscriptstyle(-)}{q}$ at the leading order QCD,
respectively.

Here we consider the following 4 pairs of the produced
hadrons $h_1$ with $z_1$ and $h_2$ with $z_2$,
$$
\rm{(i)}~~~~( \pi^+, \pi^- )~,~~~~~
\rm{(ii)}~~~~( \pi^-, \pi^+ )~,~~~~~
\rm{(iii)}~~~( \pi^+, \pi^+ )~,~~~~~
\rm{(iv)}~~~~( \pi^-, \pi^- )~,
$$
where (particle 1, particle 2) corresponds to ($h_1$ with $z_1$,
$h_2$ with $z_2$).
Then, the differential cross section of eq.(\ref{eqn:2}) for each pair
can be written as
\begin{eqnarray}
&&\rm{(i)} \nonumber\\
&&d\Delta\sigma^{\pi^+\pi^-} \nonumber\\
&&\hspace{.5cm} \sim \Delta g(\eta, Q^2)
d\Delta\hat\sigma_{PGF} \left\{ \right.
\frac{4}{9}D^{\pi^+}_u(z_1, Q^2) D^{\pi^-}_{\bar u}(z_2, Q^2)+
\frac{1}{9}D^{\pi^+}_d(z_1, Q^2) D^{\pi^-}_{\bar d}(z_2, Q^2)\nonumber\\
&&\hspace{.5cm} + \frac{1}{9}D^{\pi^+}_s(z_1, Q^2) D^{\pi^-}_{\bar s}(z_2, Q^2)+
( \pi^+(z_1)\leftrightarrow\pi^-(z_2) )\left. \right\} \nonumber\\
&&\hspace{.5cm} +  \frac{4}{9} \Delta u(\eta, Q^2) d\Delta\hat\sigma_{QCD}
\left\{ D^{\pi^+}_u(z_1, Q^2) D^{\pi^-}_g(z_2, Q^2) +
D^{\pi^-}_u(z_2, Q^2) D^{\pi^+}_g(z_1, Q^2)\right\} \nonumber\\
&&\hspace{.5cm} + ({\rm contributions~~from~~}\Delta d, \Delta s,
\Delta\bar u, \Delta\bar d~~{\rm and}~~\Delta\bar s )~,
\label{eqn:5} \\
&&\rm{(ii)} \nonumber\\
&&d\Delta\sigma^{\pi^-\pi^+} \nonumber\\
&&\hspace{.5cm} \sim \Delta g(\eta, Q^2)
d\Delta\hat\sigma_{PGF} \left\{ \right.
\frac{4}{9}D^{\pi^-}_u(z_1, Q^2) D^{\pi^+}_{\bar u}(z_2, Q^2)+
\frac{1}{9}D^{\pi^-}_d(z_1, Q^2) D^{\pi^+}_{\bar d}(z_2, Q^2)\nonumber\\
&&\hspace{.5cm} + \frac{1}{9}D^{\pi^-}_s(z_1, Q^2) D^{\pi^+}_{\bar s}(z_2, Q^2)+
( \pi^-(z_1)\leftrightarrow\pi^+(z_2) )\left. \right\} \nonumber\\
&&\hspace{.5cm} +  \frac{4}{9} \Delta u(\eta, Q^2) d\Delta\hat\sigma_{QCD}
\left\{ D^{\pi^-}_u(z_1, Q^2) D^{\pi^+}_g(z_2, Q^2) +
D^{\pi^+}_u(z_2, Q^2) D^{\pi^-}_g(z_1, Q^2)\right\} \nonumber\\
&&\hspace{.5cm} + ({\rm contributions~~from~~}\Delta d, \Delta s,
\Delta\bar u, \Delta\bar d~~{\rm and}~~\Delta\bar s )~,
\label{eqn:6} \\
&&\rm{(iii)} \nonumber\\
&&d\Delta\sigma^{\pi^+\pi^+}\nonumber\\
&&\hspace{.5cm} \sim \Delta g(\eta, Q^2)
d\Delta\hat\sigma_{PGF} \left\{ \right.
\frac{4}{9}D^{\pi^+}_u(z_1, Q^2) D^{\pi^+}_{\bar u}(z_2, Q^2)+
\frac{1}{9}D^{\pi^+}_d(z_1, Q^2) D^{\pi^+}_{\bar d}(z_2, Q^2)\nonumber\\
&&\hspace{.5cm} + \frac{1}{9}D^{\pi^+}_s(z_1, Q^2) D^{\pi^+}_{\bar s}(z_2, Q^2)+
( \pi^+(z_1)\leftrightarrow\pi^+(z_2) )\left. \right\} \nonumber\\
&&\hspace{.5cm} +  \frac{4}{9} \Delta u(\eta, Q^2) d\Delta\hat\sigma_{QCD}
\left\{ D^{\pi^+}_u(z_1, Q^2) D^{\pi^+}_g(z_2, Q^2) +
D^{\pi^+}_u(z_2, Q^2) D^{\pi^+}_g(z_1, Q^2)\right\} \nonumber\\
&&\hspace{.5cm} + ({\rm contributions~~from~~}\Delta d, \Delta s,
\Delta\bar u, \Delta\bar d~~{\rm and}~~\Delta\bar s )~,
\label{eqn:7} \\
&&\rm{(iv)} \nonumber\\
&&d\Delta\sigma^{\pi^-\pi^-}\nonumber\\
&&\hspace{.5cm} \sim \Delta g(\eta, Q^2)
d\Delta\hat\sigma_{PGF} \left\{ \right.
\frac{4}{9}D^{\pi^-}_u(z_1, Q^2) D^{\pi^-}_{\bar u}(z_2, Q^2)+
\frac{1}{9}D^{\pi^-}_d(z_1, Q^2) D^{\pi^-}_{\bar d}(z_2, Q^2)\nonumber\\
&&\hspace{.5cm} + \frac{1}{9}D^{\pi^-}_s(z_1, Q^2) D^{\pi^-}_{\bar s}(z_2, Q^2)+
( \pi^-(z_1)\leftrightarrow\pi^-(z_2) )\left. \right\} \nonumber\\
&&\hspace{.5cm} +  \frac{4}{9} \Delta u(\eta, Q^2) d\Delta\hat\sigma_{QCD}
\left\{ D^{\pi^-}_u(z_1, Q^2) D^{\pi^-}_g(z_2, Q^2) +
D^{\pi^-}_u(z_2, Q^2) D^{\pi^-}_g(z_1, Q^2)\right\} \nonumber\\
&&\hspace{.5cm} + ({\rm contributions~~from~~}\Delta d, \Delta s,
\Delta\bar u, \Delta\bar d~~{\rm and}~~\Delta\bar s )~.
\label{eqn:8}
\end{eqnarray}
Due to the isospin symmetry and charge conjugation invariance of
the fragmentation functions, various fragmentation functions in
eqs.(\ref{eqn:5})--(\ref{eqn:8})
can be classified into the following 4 functions,\cite{Kumano}
$D\equiv D^{\pi^+}_u=D^{\pi^+}_{\bar d}=D^{\pi^-}_d=D^{\pi^-}_{\bar u}$,
$\widetilde{D}\equiv D^{\pi^+}_d=D^{\pi^+}_{\bar u}=D^{\pi^-}_u
=D^{\pi^-}_{\bar d}$,
$D_s\equiv D^{\pi^+}_s=D^{\pi^+}_{\bar s}
=D^{\pi^-}_s=D^{\pi^-}_{\bar s}$ and 
$D_g\equiv D^{\pi^+}_g=D^{\pi^-}_g$,
where $D$ and $\widetilde{D}$ are called favored and unfavored
fragmentation functions, respectively.
Considering the suppression of the $s$ quark contribution to
the pion production compared with the $u$ and $d$ quark
contribution, we do not identify $D_s$ with 
$\widetilde{D}$.\cite{Kretzer,Abe}
By using these 4 kinds of pion fragmentation functions,
we can make an interesting combination of cross sections which
contains only the PGF contribution as follows;
\begin{eqnarray}
& &d\Delta\sigma^{\pi^+\pi^-}+
d\Delta\sigma^{\pi^-\pi^+}-d\Delta\sigma^{\pi^+\pi^+}-
d\Delta\sigma^{\pi^-\pi^-} 
\label{eqn:13}\\
% & &\hspace{3cm} 
& & \sim \frac{10}{9}
\Delta g(\eta, Q^2)
d\Delta\hat\sigma_{PGF}
%\times 
\left\{
D(z_1, Q^2) - \widetilde{D}(z_1, Q^2)
\right\}
\left\{
D(z_2, Q^2) - \widetilde{D}(z_2, Q^2)
\right\}~.\nonumber
\end{eqnarray}
From this combination, we can calculate the double spin
asymmetry $A_{LL}$ defined by
\begin{eqnarray}
A_{LL}&=&
\frac{d\Delta\sigma^{\pi^+\pi^-}+
d\Delta\sigma^{\pi^-\pi^+}-d\Delta\sigma^{\pi^+\pi^+}-
d\Delta\sigma^{\pi^-\pi^-}}
{d\sigma^{\pi^+\pi^-}+
d\sigma^{\pi^-\pi^+}-d\sigma^{\pi^+\pi^+}-
d\sigma^{\pi^-\pi^-}}\nonumber\\
&=&\frac{\Delta g(\eta, Q^2)}{g(\eta,
Q^2)}\cdot\frac{d\Delta\hat\sigma_{PGF}}
{d\hat\sigma_{PGF}}~,
\label{eqn:14}
\end{eqnarray}
where the factor of the fragmentation function in eq.(\ref{eqn:13})
is dropped out from the numerator and the denominator 
of $A_{LL}$.\footnote{For large $Q^2$ regions, heavy 
quarks might contribute to the PGF and QCD Compton.  
Even so, the final form of eq.(\ref{eqn:14})
for $A_{LL}$ remains unchanged
if
$D^{\pi^+}_Q=D^{\pi^+}_{\bar Q}
=D^{\pi^-}_Q=D^{\pi^-}_{\bar Q}$.}
Therefore, from the measured $A_{LL}$, one can get clear
information of $\Delta g/g$ with reliable calculation of
$d\Delta\hat\sigma_{PGF}/d\hat\sigma_{PGF}$.

\section{Numerical calculation of cross section and double spin asymmetry}
Now, let us calculate numerically the cross section and
double spin asymmetry $A_{LL}$
for the large--$p_T$ pion pair production of pol--DIS.
The spin--independent (spin--dependent) differential cross sections
for producing hadrons $h_1$ and $h_2$ are given by\cite{Peccei}
\begin{eqnarray}
\frac{d( \Delta )\sigma^{h_1 h_2}}
{dz_1d\cos\theta_1dz_2d\cos\theta_2 dx dy} &=&
\frac{d( \Delta )\sigma^{h_1 h_2}_{PGF}}
{dz_1d\cos\theta_1dz_2d\cos\theta_2 dx dy} \nonumber\\
&+&
\frac{d( \Delta )\sigma^{h_1 h_2}_{QCD}}
{dz_1d\cos\theta_1dz_2d\cos\theta_2 dx dy}~.
\label{eqn:17}
\end{eqnarray}
Each term in the right hand side of eq.(\ref{eqn:17}) is written as
\begin{eqnarray}
&&\frac{d( \Delta )\sigma^{h_1 h_2}_{PGF}}
{dz_1d\cos\theta_1dz_2d\cos\theta_2 dx dy} \nonumber\\
&&\hspace{1cm}=
(\Delta )g(\eta, Q^2)~C(\theta_1, \theta_2)
\frac{d( \Delta )\widehat{\sigma}^{h_1 h_2}_{PGF}}
{dz_id\cos\theta_1dz_{\bar i}d\cos\theta_2 dx_g dy}\nonumber\\
&&\hspace{1cm}\times \sum_{i=u,d,s,\bar u,\bar d,\bar s}
e_i^2 \{ D^{h_1}_i(\xi_1, Q^2) D^{h_2}_{\bar i}(\xi_2, Q^2) +
(1\leftrightarrow 2)\},\\
\label{eqn:18}
&&\frac{d( \Delta )\sigma^{h_1 h_2}_{QCD}}
{dz_1d\cos\theta_1dz_2d\cos\theta_2 dx dy} \nonumber\\
&&\hspace{1cm}=
\sum_{q=u, d, s, \bar u, \bar d, \bar s} e_q^2
(\Delta )f_q(\eta, Q^2)~C(\theta_1, \theta_2)
\frac{d( \Delta )\widehat{\sigma}^{h_1 h_2}_{QCD}}
{dz_qd\cos\theta_1dz_gd\cos\theta_2 dx_q dy}\nonumber\\
&&\hspace{1cm}\times \{ D^{h_1}_q(\xi_1, Q^2) D^{h_2}_g(\xi_2, Q^2) +
(1\leftrightarrow 2)\}~,
\label{eqn:19}
\end{eqnarray}
where $\eta=x+(1-x)\tau_1\tau_2$, $Q^2 = xys$, 
$\xi_1=\left (\frac{\tau_1+\tau_2}{\tau_2}\right )z_1$,
$\xi_2=\left (\frac{\tau_1+\tau_2}{\tau_1}\right )z_2$
and $C(\theta_1, \theta_2) = \frac{(\tau_1+\tau_2)^2}{\eta\tau_1\tau_2}
\frac{1-x}{8 \cos^2\frac{1}{2}\theta_1  \cos^2\frac{1}{2}\theta_2
\sin^2\frac{1}{2}(\theta_1+\theta_2) }$,
with $\tau_{1, 2}=\tan\frac{1}{2}\theta_{1, 2}$.

Here we simply assume the scattering angle of outgoing hadrons
$\theta_{1,2}$ to be the same with the one of scattered partons
in the virtual photon--nucleon c.m. frame.
This assumption might not be unreasonable if observed particles
are light hadrons with high energy.
$s$ is the total energy square of the lepton scattering off the nucleon.
$x$, $y$ and $z_{1,2}$ are
familiar kinematic variables for semi--inclusive processes in DIS,
defined as $x=\frac{Q^2}{2P\cdot q}$, $y=\frac{P\cdot q}{P\cdot\ell}$
and $z_{1,2}=\frac{P\cdot P_{1,2}}{P\cdot q}$, where $\ell$, $q$, $P$ 
and $P_{1,2}$ are the momentum of the incident lepton,
virtual photon, target nucleon and outgoing hadrons, respectively.
At the leading order of QCD, differential cross sections of 
hard scattering subprocesses, $\ell g\to \ell' q\bar q$ and 
$\ell \stackrel{\scriptscriptstyle(-)}{q}\to \ell' \stackrel{}{g}
\stackrel{\scriptscriptstyle(-)}{q}$, 
with two outgoing partons in an opposite azimuth angle
are given as\cite{DMY}
\begin{eqnarray}
\frac{d(\Delta)\hat\sigma_{PGF (QCD)}}
{dz_id\cos\theta_1d\phi_1dz_{\bar i (g)}d\cos\theta_2 dx_{g (q)}
dy d\phi_{\ell}}&=&
\frac{1}{128\pi^2}\frac{\alpha^2\alpha_s}{(p_0\cdot\ell)Q^2}
\frac{y(\eta-x)(1-\eta)^2}{x}~\nonumber\\
\times~B(\theta_1, \theta_2)~e_{\ell}^2~e_i^2|(\Delta)M|^2_{PGF (QCD)},
\label{eqn:sub}
\end{eqnarray}
with
\begin{eqnarray}
 \frac{1}{B(\theta_1,\theta_2)}=\sin(\theta_1+\theta_2)\hspace*{7cm}&& 
\nonumber\\
\times\left[\frac{\{z_i(1-\eta)+(\eta-x)\}\sin\theta_1+
\{ z_{\bar i (g)}(1-\eta)+(\eta-x) \}\sin\theta_2 }
{\sin\theta_1 \sin\theta_2}
\right],
\end{eqnarray}
where $z_i$, $z_{\bar i}$ and $z_g$ are the momentum fraction of
the outgoing parton $i$, $\bar i$ and $g$, respectively, to the incoming
parton, and are given as
$z_i=\frac{\tau_2}{\tau_1+\tau_2}$, 
$z_{\bar i (g)}=\frac{\tau_1}{\tau_1+\tau_2}$.\cite{Peccei}
The amplitude $|(\Delta)M|^2_{PGF (QCD)}$ in eq.(\ref{eqn:sub}) is
presented elsewhere\cite{Mirkes}.

By using these formulas and pion fragmentation
functions\cite{Kretzer},
we have calculated the spin--dependent and
spin--independent cross sections of the large--$p_T$ pion pair
production and estimated the double spin asymmetry 
at the energy of COMPASS experiments.\footnote{Calculation for 
HERMES energy was done\cite{DMY} without error estimation.}
Here, we have taken the AAC\cite{Goto} and GS96\cite{GS96}
parameterizations at LO QCD as polarized parton distribution functions
and GRV98\cite{GRV98} and MRST98\cite{MRST98} as unpolarized ones.
At $\sqrt s=13.7$GeV, $y=0.75$, $Q^2\geq 1$GeV$^2$ and $W^2\geq 10$GeV$^2$
with kinematical
values of $\theta_{1,2}$ and $z_{1,2}$ for
the produced pion pair, the calculated results of the spin--independent
(spin--dependent) differential cross sections and $A_{LL}$ are shown
as a function of $\eta$ in Figs.1 and 2, respectively.
In Fig.2, error was estimated by\cite{Iwata}
$\delta A_{LL}=\frac{1}{A_c P\sqrt{L}}
\sqrt{\frac{1}{d\sigma_1}+\frac{1}{d\sigma_2}}$,
where $A_c=\frac{d\sigma_1-d\sigma_2}{d\sigma_1+d\sigma_2}$
with $d\sigma_1=d\sigma(\pi^+\pi^-)+d\sigma(\pi^-\pi^+)$ and
$d\sigma_2=d\sigma(\pi^+\pi^+)+d\sigma(\pi^-\pi^-)$.
$L$ is the integrated luminosity and $P=P_B P_T f$,
where $P_B, P_T$ are the polarization of beam and target,
respectively, and $f$ is the dilution factor.
Here we used $L=2$ fb$^{-1}$(150 days running), $P_B=0.80$ and
$P_T \cdot f = 0.25$.
From Fig.2, one can see a big difference of the behavior
of $A_{LL}$ depending on the models of  $\Delta g/g$ and hence,
we can extract the behavior of $\Delta g$
rather clearly from this analysis.

\begin{figure}[htbp]
\begin{center}
%\figurebox{5pc}{2pc}{com100-2.eps} % to have a box alone
\epsfxsize=25pc % will enlarge or reduce the postscript figures based on the xsize
\epsfbox{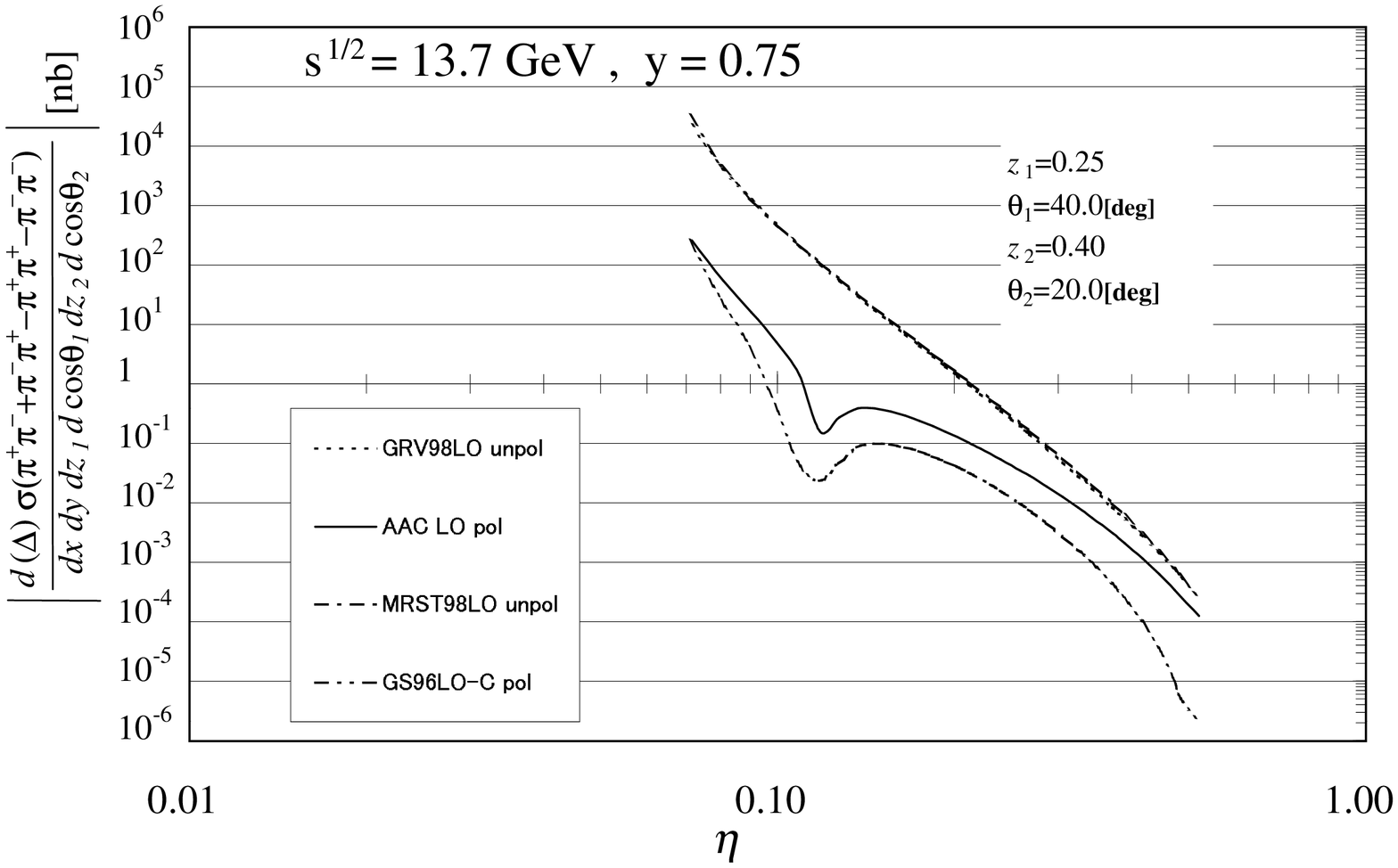} % postscript image file name
\caption{$\eta$ dependence of combined spin-independent and spin-dependent
differential cross sections defined at the denominator and numerator,
respectively, of eq.(\ref{eqn:14}) as a function of $\eta$ at
$\sqrt{s}=13.7$ GeV, $y=0.75$ for the deep inelastic regions
($Q^2 \ge 1$GeV$^2$ and $W^2 \ge 10$ GeV$^2$) with kinematical
values of $\theta_{1,2}, z_{1,2}$ for the produced pion pair.}
\end{center}
\end{figure}

\begin{figure}[htbp]
\begin{center}
%\figurebox{10pc}{5pc}{com100-1.eps}
\epsfxsize=25pc
\epsfbox{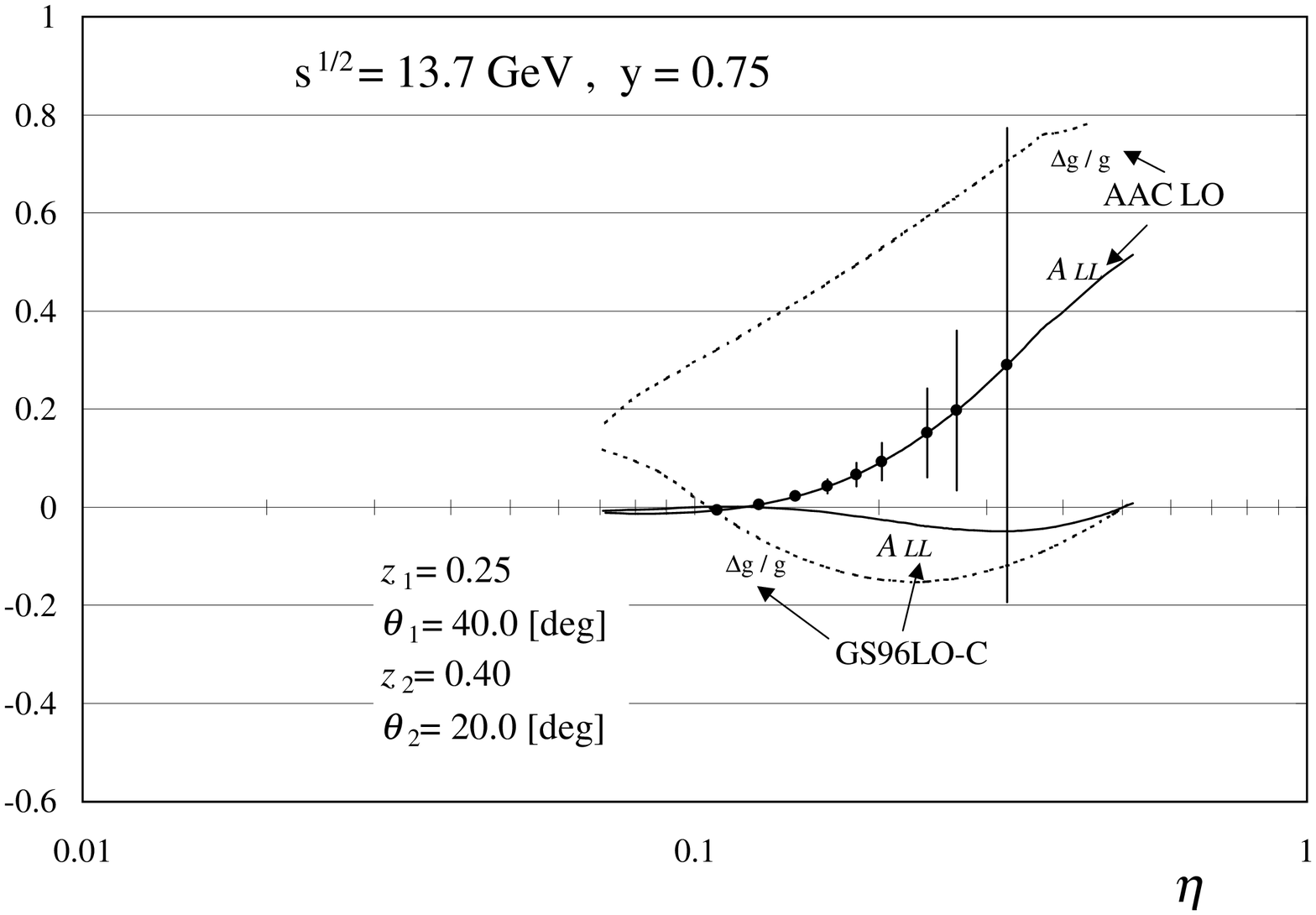}
\caption{$\eta$ dependence of $A_{LL}$ at $\sqrt{s}=13.7$ GeV, $y=0.75$
for kinematical
values of $\theta_{1,2}, z_{1,2}$ for the produced pion pair.
Solid line and dotted line are for AAC LO and GS96LO-C parameterizaton
models, respectively. $\Delta g/g$ itself is also presented
for both parameterization models.
}
\end{center}
\end{figure}

\section{Summary and discussion}

We proposed a new formula for extracting the polarized gluon distribution
from the large--$p_T$ light hadron pair production
in pol--DIS by taking an appropriate combination
of hadron pair productions.
Since the double spin asymmetry $A_{LL}$ for this combination
is directly proportional to
$\Delta g/g$, the measurement of this quantity is quite promising for
getting rather clear information on the polarized gluon
distribution in the nucleon.

The analysis can be applied also for the kaon pair or the proton pair
production by considering the reflection symmetry along the
V--spin axis, the isospin symmetry and charge conjugation invariance of
the fragmentation functions.

The same combinations of cross sections for light hadron pair
production were discussed for photoproduction
by other people\cite{GCV},
while we studied the lepton-proton processes in deep inelastic region
in this work. Both of those processes are useful for extracting $\Delta g$.

\vspace{1em}
This work is supported by the Grant--in--Aid for Science Research,
Ministry of Education, Science and Culture, Japan (No.11694081).

\end{document}